 \documentclass[final,5p,times,twocolumn]{elsarticle}

\usepackage{bm}
\usepackage{amsmath}
\usepackage{xcolor}
\usepackage{amsmath}
\usepackage[abs]{overpic}
\usepackage{array}
\makeatletter
\renewcommand*\env@matrix[1][\arraystretch]{%
  \edef\arraystretch{#1}%
  \hskip -\arraycolsep
  \let\@ifnextchar\new@ifnextchar
  \array{*\c@MaxMatrixCols c}}
\makeatother 
\biboptions{sort&compress}

\usepackage{amssymb}
\journal{Physics Letters B}

\definecolor{c1}{HTML}{000000}
\definecolor{c2}{HTML}{00FF00}
\definecolor{c3}{HTML}{000000}
\definecolor{c4}{HTML}{00FFFF}
\definecolor{c5}{HTML}{DF8500}
\renewcommand{\eqref}[1]{(\ref{#1})}
\newcommand{\Sec}[1]{Sec.~\ref{#1}}
\newcommand{\Fig}[1]{Fig.~\ref{#1}}
\newcommand{\Eq}[1]{Eq.~(\ref{#1})}

\newcommand{\Ref}[1]{Ref.~\cite{#1}}
\newcommand{\Refs}[1]{Refs.~\cite{#1}}
\newcommand{\Tab}[1]{Table~\ref{#1}}

\begin{document}

\begin{frontmatter}



\title{${^3{\rm He}}(\alpha,\gamma){^7{\rm Be}}$ and ${^3{\rm H}}(\alpha,\gamma){^7{\rm Li}}$ astrophysical $S$ factors from the no-core shell model with continuum}


\author[TRIUMF]{J\'er\'emy Dohet-Eraly}
\ead{jdoheter@triumf.ca}

\author[TRIUMF]{Petr Navr\'atil}
\ead{navratil@triumf.ca}

\author[LLNL]{Sofia Quaglioni}
\ead{quaglioni1@llnl.gov}

\author[Hokk]{Wataru Horiuchi}
\ead{whoriuchi@nucl.sci.hokudai.ac.jp}

\author[LLNL,Orsay]{Guillaume Hupin\fnref{new}}
\ead{hupin@ipno.in2p3.fr}
\fntext[new]{Present address: CEA, DAM, DIF, F-91297 Arpajon, France.}

\author[TRIUMF]{Francesco Raimondi\fnref{new2}}
\ead{f.raimondi@surrey.ac.uk}
\fntext[new2]{Present address: Department of Physics, Faculty of Engineering and Physical Sciences,
University of Surrey, Guildford, Surrey GU2 7XH, United Kingdom.}

\address[TRIUMF]{TRIUMF, 4004 Wesbrook Mall, Vancouver BC V6T 2A3, Canada}
\address[LLNL]{Lawrence Livermore National Laboratory, P.O. Box 808, L-414, Livermore, California 94551, USA}
\address[Hokk]{Department of Physics, Hokkaido University, Sapporo 060-0810, Japan}
\address[Orsay]{Institut de Physique Nucl\'eaire, IN2P3-CNRS, Universit\'e Paris-Sud, F-91406 Orsay  Cedex, France.}

\begin{abstract}
The ${^3{\rm He}}(\alpha,\gamma){^7{\rm Be}}$ and ${^3{\rm H}}(\alpha,\gamma){^7{\rm Li}}$ astrophysical $S$ factors are calculated 
within the no-core shell model with continuum using a renormalized chiral nucleon-nucleon interaction. 
The  ${^3{\rm He}}(\alpha,\gamma){^7{\rm Be}}$ astrophysical $S$ factors agree reasonably well with the experimental data while the ${^3{\rm H}}(\alpha,\gamma){^7{\rm Li}}$ ones are overestimated.
The seven-nucleon bound and resonance states and  the $\alpha+{^3{\rm He}}/{^3{\rm H}}$ elastic scattering are also studied and compared with experiment. The low-lying resonance properties are rather well reproduced by our approach. At low energies, the $s$-wave phase shift, which is non-resonant, is overestimated.
\end{abstract}


\end{frontmatter}


\hyphenation{NCSMC}
\hyphenation{NCSM}
\hyphenation{RGM}
\section{Introduction}
The ${^3{\rm He}}(\alpha,\gamma){^7{\rm Be}}$ and ${^3{\rm H}}(\alpha,\gamma){^7{\rm Li}}$ radiative-capture processes hold great astrophysical significance. 
Their reaction rates for collision energies between  $\sim$20 and 500~keV  in the center-of-mass (c.m.)\ frame are essential to calculate the primordial ${^7{\rm Li}}$ abundance in the universe~\cite{BNT99,NB00,No01}. 
In addition, standard solar model predictions for the fraction of pp-chain branches resulting in ${^7{\rm Be}}$ versus ${^8{\rm B}}$ neutrinos depend critically on the ${^3{\rm He}}(\alpha,\gamma){^7{\rm Be}}$ astrophysical $S$ factor at about 20~keV c.m.\ energy~\cite{AAB98,AGR11}. 
Because of the Coulomb repulsion between the fusing nuclei, these capture cross sections are strongly suppressed at such low energies and thus hard to measure directly in a laboratory.

Concerning the ${^3{\rm He}}(\alpha,\gamma){^7{\rm Be}}$ radiative capture, experiments performed by several groups in the last decade have lead to quite accurate cross-section determinations for collision energies between about 90~keV and 3.1~MeV in the c.m.\ frame~\cite{NHN04,BCC06,CBC07,BBS07,DGK09,CNB12,BGH13,Ca14}. 
However, theoretical models or extrapolations are still needed to provide the capture cross section at solar energies~\cite{DGS16}.
In contrast, experimental data are much less accurate for the ${^3{\rm H}}(\alpha,\gamma){^7{\rm Li}}$ radiative capture.
The most recent experiment was performed twenty years ago resulting in measurements at collision energies between about 50~keV and 1.2~MeV in the c.m.\ frame~\cite{BKR94}.

Theoretically, these radiative captures have also generated much interest: from the development of pure external-capture models in the early 60's~\cite{TP63} to the microscopic approaches from the late 80's up to now~\cite{Ka86,MH86,CL00,No01,Ne11} (see \Ref{AGR11} for a short review). 
However, no parameter-free approach is able to simultaneously reproduce the latest experimental ${^3{\rm He}}(\alpha,\gamma){^7{\rm Be}}$ and ${^3{\rm H}}(\alpha,\gamma){^7{\rm Li}}$ astrophysical $S$ factors.
To possibly fill this gap, an \textit{ab initio} approach, relying on a realistic inter-nucleon interaction, is highly desirable.
The \textit{ab initio} no-core shell model with continuum (NCSMC)~\cite{BNQ13,BNQ13b} has been successful in the simultaneous description of bound and scattering states associated with realistic Hamiltonians~\cite{HQN14,HQN15}.
This approach can thus be naturally applied to the description of radiative-capture reactions, which involve both scattering (in the initial channels) and bound states (in the final channels).

In this letter, we present the study of the ${^3{\rm He}}(\alpha,\gamma){^7{\rm Be}}$ and ${^3{\rm H}}(\alpha,\gamma){^7{\rm Li}}$ radiative-capture reactions with the NCSMC approach~\cite{BNQ13,BNQ13b}, using a renormalized chiral nucleon-nucleon ($NN$) interaction.
This is the first NCSMC study where the lightest colliding nucleus has three nucleons and the first application of the NCSMC to a radiative capture. 
We outline the NCSMC formalism in \Sec{Secform} and apply the NCSMC approach to the study of seven-nucleon systems in \Sec{Secres}. First, properties of ${^7{\rm Be}}$ and ${^7{\rm Li}}$ bound states and resonance states are evaluated and compared with the experimental data. 
Then, the $\alpha+{^3{\rm He}}$ and $\alpha+{^3{\rm H}}$ scattering states are studied. 
Elastic cross sections, elastic phase shifts, and scattering lengths are computed and compared with experimental data or values obtained by other models. Finally, from the ${^7{\rm Be}}$ and ${^7{\rm Li}}$ bound-state wave functions and the $\alpha+{^3{\rm He}}$ and $\alpha+{^3{\rm H}}$ scattering wave functions, we evaluate the ${^3{\rm He}}(\alpha,\gamma){^7{\rm Be}}$ and ${^3{\rm H}}(\alpha,\gamma){^7{\rm Li}}$ radiative-capture cross sections.
\section{Formalism}\label{Secform}
The present study of the ${^7{\rm Be}}$ and ${^7{\rm Li}}$ nuclei, of the $\alpha+{^3{\rm He}}$ and $\alpha+{^3{\rm H}}$ elastic scattering, and of the  ${^3{\rm He}}(\alpha,\gamma){^7{\rm Be}}$ and ${^3{\rm H}}(\alpha,\gamma){^7{\rm Li}}$ radiative-capture reactions is based on the solutions of the microscopic Schr\"odinger equation
\begin{equation}\label{Sch}
H |\Psi^{J^\pi T}\rangle=\left[\sum^7_{i=1} t_i -T_{\rm c.m.} + \sum^7_{i<j=1} v_{ij}\right] |\Psi^{J^\pi T}\rangle=E |\Psi^{J^\pi T}\rangle,
\end{equation}
at different energies for different values of angular momentum $J$, parity $\pi$, and isospin $T$. 
The quantum numbers associated with the projections of the angular momentum and of the isospin are omitted for simplifying the notations.
Here $H$ is the translation-invariant microscopic Hamiltonian, $t_i$ is the kinetic energy of nucleon $i$, $T_{\rm c.m.}$ is the c.m.\ kinetic energy, $v_{ij}$ is the potential between nucleons $i$ and $j$, $E$ is the total energy in the c.m.\ frame, and  $|\Psi^{J^\pi T}\rangle$ is a partial wave function with quantum numbers ${J^\pi T}$.
The $NN$ potential $v_{ij}$, which is specified in the next section, is realistic, in the sense that it reproduces the experimental deuteron energy and $NN$ phase shifts. For computational reasons, no three-body forces are considered in the present work.

For the sake of brevity, in the rest of this section, the formalism is presented only for the $^7{\rm Be}$/$\alpha+{^3{\rm He}}$ system.
The treatment of the $^7{\rm Li}$/$\alpha+{^3{\rm H}}$ system is analogous.
In the NCSMC approach, the colliding nuclei, $\alpha$ and ${^3{\rm He}}$, are described by square-integrable eigenstates of the ${^4{\rm He}}$ and ${^3{\rm He}}$ systems obtained within the no-core shell model (NCSM)~\cite{NVB00a,NQS09} by diagonalizing a large matrix. These NCSM eigenstates are linear combinations of translation-invariant and fully-antisymmetric states of harmonic oscillator (HO) wave functions with frequency $\Omega$ and up to $N_{\rm max}$ HO quanta above the lowest energy configuration. 
The compound system, the $^7{\rm Be}$ bound and resonance states and the $\alpha+{^3{\rm He}}$ scattering states, are described by a combination of seven-body NCSM eigenstates, denoted by $|7\lambda J^\pi T\rangle$ with $\lambda$ the energy label,  and ${^4{\rm He}}+{^3{\rm He}}$ cluster states, denoted by $|\Phi^{J^\pi T}_{\nu r}\rangle$, and built from the NCSM eigenstates of ${^4{\rm He}}$ and ${^3{\rm He}}$ by means of the resonating-group method (RGM)~\cite{QN08,QN09}.
The latter cluster states are explicitly given by 
\begin{equation}
\begin{split}
|\Phi^{J^\pi T}_{\nu r}\rangle=&\left[\left[|{^4{\rm He}} \lambda_4 J^{\pi_4}_4  T_4\rangle
|{^3{\rm He}} \lambda_3 J^{\pi_3}_3 T_3\rangle\right]^{(sT)} Y_\ell(\hat{r}_{43})\right]^{(J^\pi T)}\\
&\displaystyle \times \frac{\delta(r-r_{43})}{r r_{43}},
\end{split}
\end{equation}
where $|{^4{\rm He}} \lambda_4 J^{\pi_4}_4  T_4\rangle$ is a NCSM eigenstate of ${^4{\rm He}}$ with energy label $\lambda_4$, total angular momentum $J_4$, parity $\pi_4$, and isospin $T_4$, $|{^3{\rm He}} \lambda_3 J^{\pi_3}_3  T_3\rangle$ is an analogously defined NCSM eigenstate of ${^3{\rm He}}$, $s$ is the channel spin, $\bm{r}_{43}$ is the relative coordinate between the centers of mass of ${^4{\rm He}}$ and ${^3{\rm He}}$, $\ell$ is the relative orbital angular momentum between the clusters, and $\nu$ is a collective index for $\{\lambda_4, J_4, \pi_4, T_4,\lambda_3, J_3,\pi_3,  T_3,s,\ell\}$. 
These cluster states are translation-invariant. Full antisymmetrization is obtained by applying an operator $\hat{\mathcal{A}}_{43}$ that accounts for exchanges between the nucleons belonging to ${^4{\rm He}}$ and those belonging to ${^3{\rm He}}$.
The NCSMC partial wave function is thus expanded as
\begin{equation}\label{NCSMCwf}
\begin{split}
|\Psi^{J^\pi T}\rangle=&\sum_\lambda c^{J^\pi T}_\lambda |7\lambda J^\pi T\rangle\\
&+\sum_\nu \int dr r^2 \frac{\gamma^{J^\pi T}_\nu(r)}{r} \hat{\mathcal{A}}_{43} |\Phi^{J^\pi T}_{\nu r}\rangle,
\end{split}
\end{equation}
where the coefficients $c^{J^\pi T}_\lambda$ and functions $\gamma^{J^\pi T}_\nu$ are unknown discrete and continuous amplitudes.
 
Inserting ansatz~\eqref{NCSMCwf} in \Eq{Sch} and projecting over the basis states lead to the NCSMC equations~\cite{BNQ13b}, written in a schematic block-matrices notation as
\begin{equation}\label{NCSMCeq}
\begin{pmatrix}[1.3] H^{J^\pi T}_{7} & h^{J^\pi T}\\h^{J^\pi T}& \mathcal{H}^{J^\pi T} \end{pmatrix} \begin{pmatrix}[1.3]c^{J^\pi T}\\\gamma^{J^\pi T}  \end{pmatrix}=
E \begin{pmatrix}[1.3] I^{J^\pi T}_{7} & g^{J^\pi T}\\g^{J^\pi T}& \mathcal{N}^{J^\pi T} \end{pmatrix} \begin{pmatrix}[1.3]c^{J^\pi T}\\\gamma^{J^\pi T}  \end{pmatrix}.
\end{equation}
The upper diagonal blocks in the left- and right-hand sides of the equation are the Hamiltonian and overlap matrix elements on the square-integrable seven-body states obtained from the NCSM diagonalization, 
\begin{eqnarray}\label{Elam}
(H^{J^\pi T}_7)_{\lambda \lambda'}&=&\langle 7\lambda J^\pi T|H|7\lambda' J^\pi T\rangle=E_\lambda \delta_{\lambda \lambda'},\\
(I^{J^\pi T}_7)_{\lambda \lambda'}&=&\langle 7\lambda J^\pi T|7\lambda' J^\pi T\rangle=\delta_{\lambda \lambda'}.
\end{eqnarray}
The lower diagonal blocks are the Hamiltonian and overlap matrix elements on the ${^4{\rm He}}+{^3{\rm He}}$ cluster states,
\begin{eqnarray}
\mathcal{H}^{J^\pi T}_{\nu \nu'}(r,r') &=& \langle \Phi^{J^\pi T}_{\nu r}|\hat{\mathcal{A}}_{43} H\hat{\mathcal{A}}_{43} |\Phi^{J^\pi T}_{\nu' r'}\rangle, \\
\mathcal{N}^{J^\pi T}_{\nu \nu'}(r,r') &=& \langle \Phi^{J^\pi T}_{\nu r}| \hat{\mathcal{A}}^2_{43} |\Phi^{J^\pi T}_{\nu' r'}\rangle.
\end{eqnarray}
The off-diagonal blocks are the coupling Hamiltonian and overlap matrix elements between square-integrable states and cluster states,
\begin{eqnarray}
h^{J^\pi T}_{\lambda \nu}(r)&=&\langle 7\lambda J^\pi T|H\hat{\mathcal{A}}_{43} |\Phi^{J^\pi T}_{\nu r}\rangle,\\
g^{J^\pi T}_{\lambda \nu}(r)&=&\langle 7\lambda J^\pi T|\hat{\mathcal{A}}_{43} |\Phi^{J^\pi T}_{\nu r}\rangle.
\end{eqnarray}
The system of coupled equations~\eqref{NCSMCeq}, transformed first into an orthogonal form~\cite{BNQ13b}, is solved by means of the microscopic $R$-matrix method on a Lagrange mesh~\cite{BHL77,HSV98,DB10}, which enforces the proper bound-state or scattering-state asymptotic behavior of functions $\gamma^{J^\pi T}_\nu$.
The asymptotic normalization coefficients and the phase shifts (or the collision matrix elements if several channels are open) are then computed from the $R$-matrix.
When only one channel is open, the value of the $R$-matrix at zero colliding energy enables a simple and accurate evaluation of the scattering lengths~\cite{BHK00}.
Resonances can also be studied with this approach. 
Indeed, extending the microscopic $R$-matrix approach to complex energies~\cite{CH97} allows the determination of the Siegert states~\cite{Si39}, which are solutions of the Schr\"odinger equation~\eqref{Sch} with purely outgoing waves at infinite relative distance $r_{43}$.
The complex energies of these states are the poles of the $S$-matrix, thus providing directly the energy and width of the resonances.

Finally, by computing the matrix elements of the electromagnetic multipole operators between the initial $\alpha+{^3{\rm He}}$ scattering state, corresponding to the energy of the collision, and the final ${^7{\rm Be}}$ bound state, the ${^3{\rm He}}(\alpha,\gamma){^7{\rm Be}}$ astrophysical $S$ factors can be evaluated~\cite{BD83}.
\section{Results}\label{Secres}
The nucleon-nucleon interaction is described by a chiral N$^3$LO $NN$ potential~\cite{EM03} softened via the similarity-renormalization-group (SRG) method~\cite{We94,BFP07,JNF09,JNF11}, which reduces the influence of momenta higher than an SRG resolution scale $\hbar \Lambda$.
For computational reasons, in the present applications, both chiral and renormalization-induced three- and higher-body forces are disregarded.
To get the correct tail of the bound-state wave functions and hence sensible astrophysical $S$ factors, the SRG resolution scale is chosen as $\Lambda=2.15~{\rm fm}^{-1}$ to reproduce, as accurately as possible, the experimental separation energies of the ${^7{\rm Be}}$ and ${^7{\rm Li}}$ nuclei for the largest accessible model space (see below for its specifications).
An analogous strategy has already been followed in a similar context: the study of the ${^7{\rm Be}}(p,\gamma){^8{\rm B}}$ radiative capture with the NCSM/RGM approach~\cite{NRQ11}.

\begin{figure}[ht]
\centering
 \scalebox{0.57}{\includegraphics{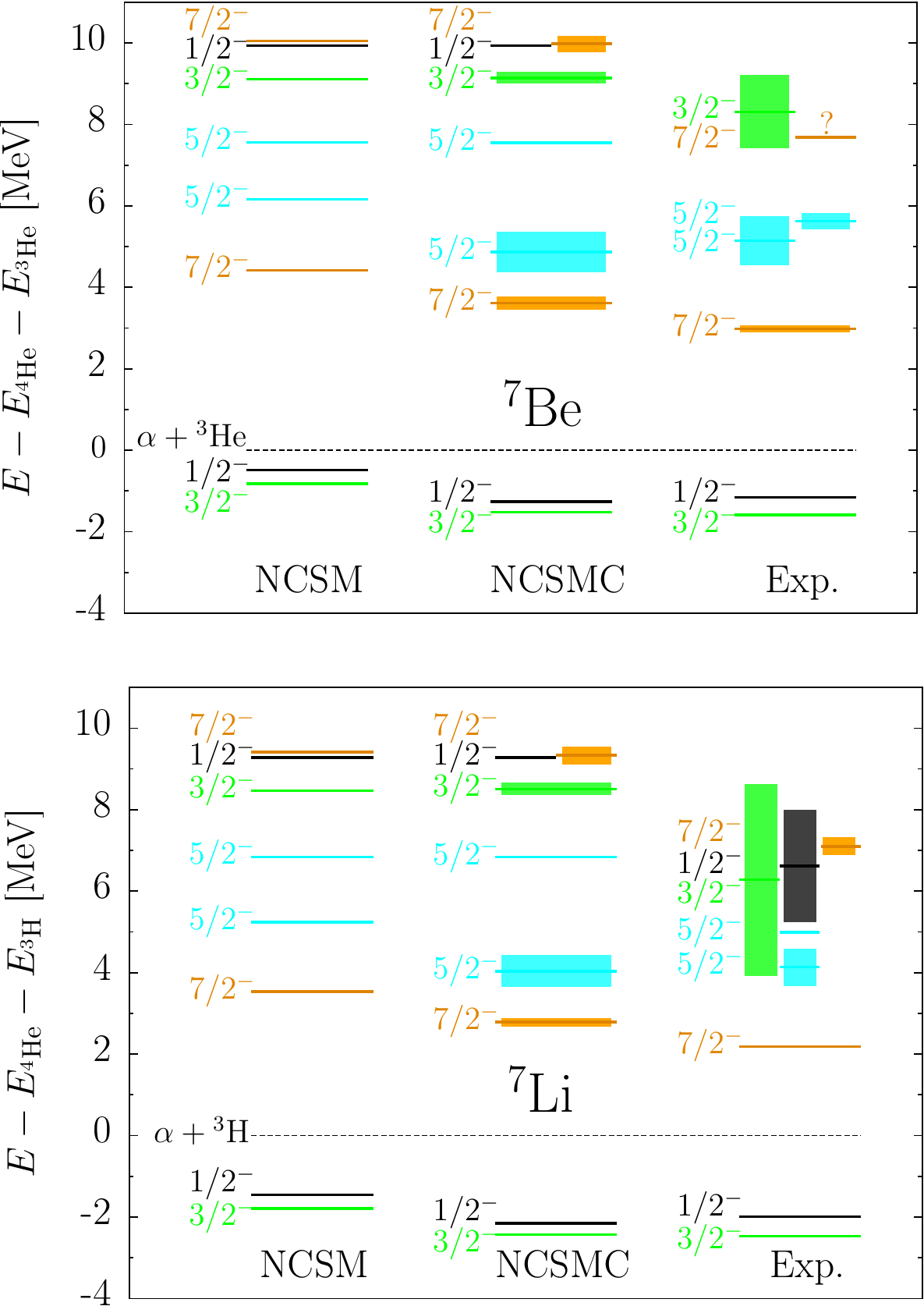}}
\caption{(Color online) The $^7{\rm Be}$ and $^7{\rm Li}$  spectra obtained from the NCSM and NCSMC approaches and from experiments~\cite{TCG02}. Only states with isospin $T=1/2$ are considered. 
Energies are given with respect to the $\alpha$+${^3{\rm He}}$/${^3{\rm H}}$ threshold. Rectangles symbolize the widths of resonances. The question mark indicates that the width is not experimentally determined.}
\label{spectra}
 \end{figure}

All partial waves with total angular momentum $J\in\{1/2,3/2,5/2,7/2\}$, positive or negative parities, and isospin $T=1/2$ are considered. The HO frequency is $\Omega=20$~MeV$/\hbar$ for both square-integrable states and cluster states included in expansion~\eqref{NCSMCwf}.
All NCSM eigenstates up to about $13$~MeV with respect to the $\alpha+^3{\rm He}$/$^3{\rm H}$ threshold are included in the model space, which corresponds to one or two states by partial wave. As an example, the negative-parity NCSM states can be seen in the leftmost column of \Fig{spectra}.
The values $N_{\rm max}=11$ for positive-parity states and $N_{\rm max}=10$ for negative-parity states are considered. 
The ${^4{\rm He}}+{^3{\rm He}}$ (${^4{\rm He}}+{^3{\rm H}}$) cluster states are built by coupling the NCSM ground states of ${^4{\rm He}}$ $[(J^\pi T)=(0^+0)]$ and ${^3{\rm He}}$ (${^3{\rm H}}$) $[(J^\pi T)=(1/2^+ 1/2)]$ obtained with $N_{\rm max}=12$, except if another value is explicitly specified. 
For studying  the convergence properties, smaller values of $N_{\rm max}$ are used but the gap between the values of $N_{\rm max}$ used for computing the colliding-nuclei wave functions, the seven-body positive-parity states and the seven-body negative-parity states is always the same.
No cluster states involving excited states of ${^4{\rm He}}$ or ${^3{\rm He}}$ (${^3{\rm H}}$) are considered.
The orthogonal version of NCSMC equations \eqref{NCSMCeq} are solved by the microscopic $R$-matrix with a channel radius $a=12$~fm and $40$ Lagrange-mesh points.

The ${^4{\rm He}}$, ${^3{\rm He}}$, and ${^3{\rm H}}$ ground-state energies and charge radii obtained with $N_{\rm max}=12$ are given in \Tab{tab0}. They are compared with the {\it exact} values obtained by increasing $N_{\rm max}$ up to convergence and with the experimental data. 
\begin{table}[ht]
\centering
\renewcommand{\arraystretch}{1.2}
\begin{tabular}{l l l l }
\hline
Nucleus&$E_{\rm g.s.}$ [MeV]&$r_{\rm ch}$ [fm]& \\
\hline
 ${^4{\rm He}}$ &-27.97 &1.68 & $N_{\rm max}=12$\\
 &-28.03 &1.68 & converged \\
 &-28.296 &1.681(4) &Exp.~\cite{TWH92,Si14}\\
 ${^3{\rm He}}$ &-7.49 &1.95 &$N_{\rm max}=12$\\
 &-7.55 &2.00 &converged  \\
 &-7.718 &1.973(14) &Exp.~\cite{PKK10,Si14}\\
 ${^3{\rm H}}$ &-8.24 &1.76 &$N_{\rm max}=12$\\
 &-8.30 &1.79 & converged \\
 &-8.482& 1.7591(363) &Exp.~\cite{PKK10,AM13}\\
\hline
\end{tabular}
\caption{Ground-state energies and charge radii of the ${^4{\rm He}}$, ${^3{\rm He}}$, and ${^3{\rm H}}$ nuclei calculated with the no-core shell model with $N_{\rm max}=12$ and compared with converged values (obtained with higher values of $N_{\rm max}$) and with experimetal data (\Refs{TWH92,PKK10} for energies and \Refs{Si14,AM13} for radii). The chiral N$^3$LO $NN$ potential softened via the SRG with $\Lambda=2.15~{\rm fm}^{-1}$ is used.
}
\label{tab0}
\renewcommand{\arraystretch}{1}
\end{table}
At $N_{\rm max}=12$, the ${^4{\rm He}}$, ${^3{\rm He}}$, and ${^3{\rm H}}$ ground-state properties are close to their converged values with a relative difference of less than $1\%$ for the energies and less than $3\%$ for the radii. They are also close to the experimental values: the energies at $N_{\rm max}=12$ differ from the experimental ones by less than $3\%$ while the radii at  $N_{\rm max}=12$ differ by less than $1\%$ from the experimental ones or are in agreement with them.

In the following, we discuss the $^7{\rm Be}$ and $^7{\rm Li}$ bound-state properties as they are obtained within the NCSMC approach. 
Both nuclei have a ground state characterized by $(J^\pi T)=(3/2^- 1/2)$ and an excited state characterized by $(J^\pi T)=(1/2^- 1/2)$. 
Their energies are displayed in \Tab{tab1} and compared with the values obtained with the square-integrable part of the basis only and with the experimental values. The charge radii ($r_{\rm ch}$), quadrupole moments ($Q$), and magnetic dipole moments ($\mu$) of the $^7{\rm Be}$ and $^7{\rm Li}$ ground states are also given in  \Tab{tab1}.

\begin{table}[ht]
\centering
\renewcommand{\arraystretch}{1.2}
\begin{tabular}{p{0.4cm} l r  r l  l}
\hline
$^7{\rm Be}$&&\multicolumn{1}{c}{NCSM}&\multicolumn{1}{c}{NCSMC}&\multicolumn{1}{c}{Exp.}&Refs.\\
\hline
$E_{3/2 ^-}$&[MeV]& -0.82  &-1.52&-1.587&\cite{TCG02}\\
$E_{1/2 ^-}$&[MeV]&-0.49 &-1.26&-1.157&\cite{TCG02}\\
$r_{\rm ch}$&[fm]& 2.375 &2.62&2.647(17)&\cite{NTZ09}\\
$Q$&${\rm [e\ fm^2]}$&-4.57&-6.14&-&\\
$\mu$&$[\mu_N]$&-1.14&-1.16&-1.3995(5)&\cite{NTZ09}\\
\hline
$^7{\rm Li}$&&\multicolumn{1}{c}{NCSM}&\multicolumn{1}{c}{NCSMC}&\multicolumn{1}{c}{Exp.}&Refs.\\
\hline
$E_{3/2 ^-}$&[MeV]&  -1.79 & -2.43&-2.467&\cite{TCG02}\\
$E_{1/2 ^-}$&[MeV]&-1.46&-2.15&-1.989&\cite{TCG02}\\
$r_{\rm ch}$&[fm]&2.21&2.42&2.39(3)&\cite{DDD74}\\
$Q$&${\rm [e\ fm^2]}$&-2.67&-3.72&-4.00(3)&\cite{VF91}\\
$\mu$&$[\mu_N]$&3.00&3.02&3.256&\cite{Ra89}\\
\hline
\end{tabular}
\caption{Properties of the $^7{\rm Be}$ and $^7{\rm Li}$ bound states calculated within the NCSM and NCSMC approaches and compared with experimental data~\cite{TCG02,NTZ09,DDD74,VF91,Ra89}. 
%Parameters of calculations are given in the text.
The $^7{\rm Be}$ and $^7{\rm Li}$ energies are given with respect to the $\alpha$+${^3{\rm He}}$ and $\alpha$+${^3{\rm H}}$ thresholds, respectively.}
\label{tab1}
\renewcommand{\arraystretch}{1}
\end{table}
Comparing NCSM and NCSMC results shows that the explicit inclusion of cluster states (for a given $N_{\rm max}$) has a strong impact on the energies, charge radii, and quadrupole moments.
For all these quantities but the $^7{\rm Be}$ quadrupole moment, which has not been measured yet, the inclusion of cluster basis states significantly reduces the gap between calculated and experimental values.
On the contrary, the magnetic dipole moment is little affected by the presence of the $\alpha+{^3{\rm He}}/{^3{\rm H}}$ cluster degrees of freedom. 
The discrepancy between its theoretical and experimental values is mostly due to the two-body electromagnetic currents, which are missing in our approach.

An analogous comparison for the $^7{\rm Be}$ and $^7{\rm Li}$ computed and measured spectra is given in \Tab{tab2} and in \Fig{spectra}.
\begin{table}[h!]
\centering
\renewcommand{\arraystretch}{1.2}
{\small
\begin{tabular}{l r r r r r}
\hline
$^7{\rm Be}$&\multicolumn{1}{c}{NCSM}&\multicolumn{2}{c}{NCSMC}&\multicolumn{2}{c}{Exp.}\\
\hline
$J^\pi$ &\multicolumn{1}{c}{$E_r$}& \multicolumn{1}{c}{$E_r$} &\multicolumn{1}{c}{$\Gamma$}&  \multicolumn{1}{c}{$E_r$} &\multicolumn{1}{c}{$\Gamma$}\\
$7/2^-$& 4.42&3.61 & 0.33                          &   2.98 &  0.175\\
$5/2^-$& 6.16&4.87 & 1.00                          &   5.14&  1.2\\
$5/2^-$& 7.56&7.55 & $\ast$& 5.62& 0.40\\
$3/2^-$& 9.11&9.14 & 0.29                          &   8.31&  1.8\\
$1/2^-$& 9.93&9.93 & $\ast$        &-      &-\\
$7/2^-$& 10.05&9.98 & 0.40                         & 7.68&  -\\
\hline
$^7{\rm Li}$&\multicolumn{1}{c}{NCSM}&\multicolumn{2}{c}{NCSMC}&\multicolumn{2}{c}{Exp.}\\
\hline
$J^\pi$ & \multicolumn{1}{c}{$E_r$} & \multicolumn{1}{c}{$E_r$} & \multicolumn{1}{c}{$\Gamma$}&  \multicolumn{1}{c}{$E_r$} & \multicolumn{1}{c}{$\Gamma$}\\
$7/2^-$&3.53& 2.79 & 0.214&                                    2.18 &  0.069\\
$5/2^-$&5.24& 4.04 & 0.785&                                    4.14&  0.918\\
$5/2^-$&6.84& 6.84 & $\ast$&        4.99 &  0.080\\
$3/2^-$&8.47& 8.51 & 0.297&                                    6.28 &  4.712\\
$1/2^-$&9.28& 9.28 & $\ast$&       6.62  &  2.752\\
$7/2^-$&9.41& 9.33 & 0.435&                                   7.10   &  0.437\\
\hline
\end{tabular}
}
\caption{Energies ($E_r$) and widths ($\Gamma$) in MeV of $^7{\rm Be}$ and $^7{\rm Li}$ resonance states up to $10$~{\rm MeV} obtained from the NCSM and NCSMC approaches and from experiments~\cite{TCG02}. 
Only resonances with isospin $T=1/2$ are considered. 
Resonance energies are given with respect to the $\alpha$+${^3{\rm He}}/{^3{\rm H}}$ threshold. Experimental uncertainties are not displayed. Widths marked with an asterisk cannot be extracted reliably because the ${^6{\rm Li}}+N$ decay channel is not considered in these calculations.}
\label{tab2}
\renewcommand{\arraystretch}{1}
\end{table}
Note that the NCSM approach can only provide the energies of resonances, not their widths. 
Including the $\alpha+{^3{\rm He}}$ ($\alpha+{^3{\rm H}}$) cluster states in the model space reduces the gap between the theoretical and experimental energies of the first two resonances while the energies of the other resonances are nearly unaffected.
The first $7/2^-$ resonance is overestimated by 600~keV while the first $5/2^-$ resonance is underestimated by 300~keV for the ${^7{\rm Be}}$ system and by 100~keV for the ${^7{\rm Li}}$ system.
The widths of the second $5/2^-$ and of the $1/2^-$ resonances obtained with the NCSMC approach are unphysically small (less than 10~keV). The escape width for these resonances is missing because the corresponding decay channel (${^6{\rm Li}}+N$) is not included in the calculations. 
The explicit inclusion of ${^6{\rm Li}}+N$ cluster states in the basis should cure this problem and also affect the other states close or above the ${^6{\rm Li}}+N$ threshold.
While at energies relevant for astrophysics the radiative capture is clearly non-resonant, the first $7/2^-$ resonance in the ${^7{\rm Be}}$ spectrum plays a minor but non-negligible role at the relevant energies for laboratory measurements, as it is shown further. 

\begin{figure}
\centering
 \scalebox{0.3}{\includegraphics{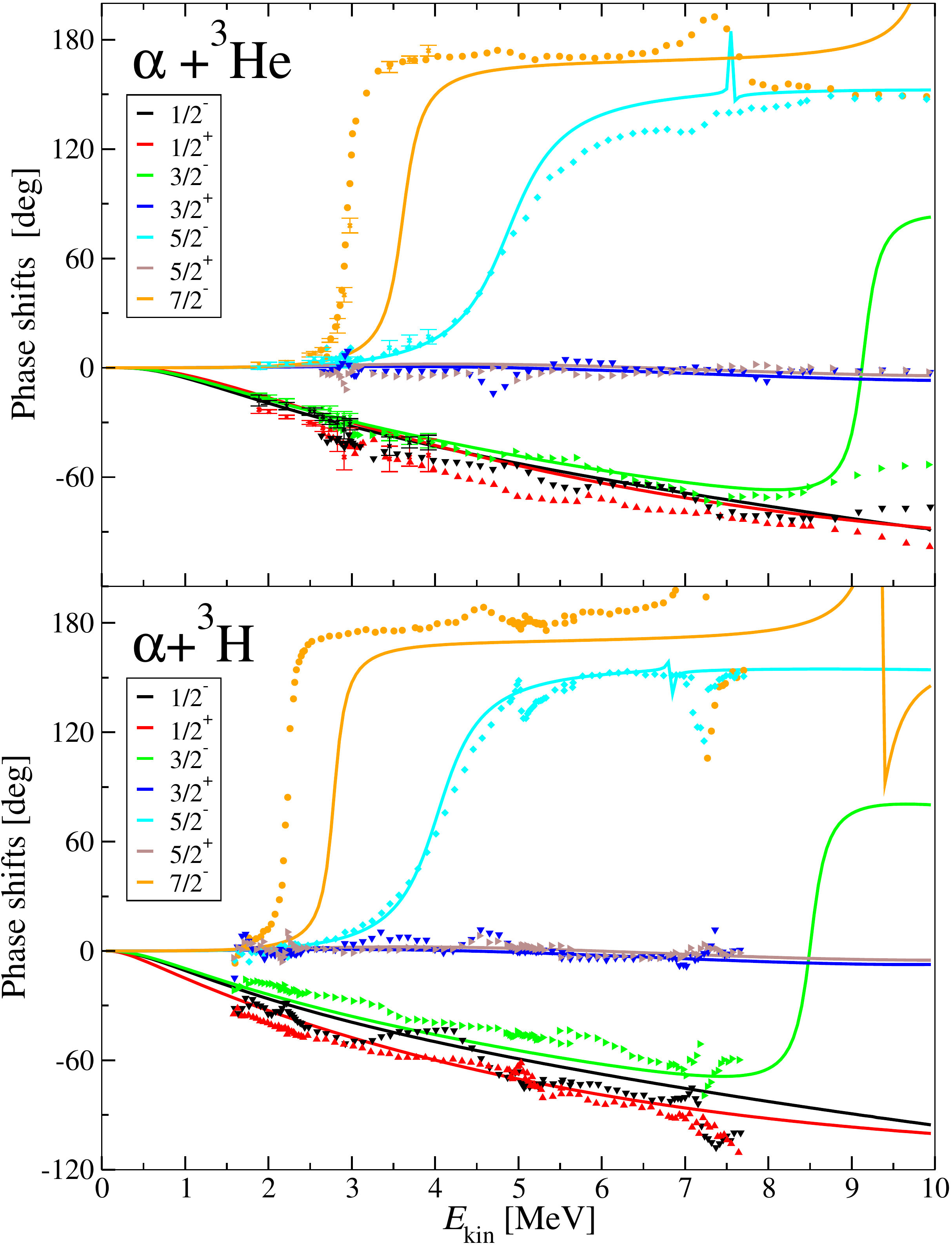}}
% \scalebox{0.3}{\includegraphics{phase_shift_3H3He4He}}
\caption{(Color online) The $\alpha+{^3{\rm He}}$ and $\alpha+{^3{\rm H}}$ elastic phase shifts obtained from the NCSMC approach and from experiments~\cite{ST67,BBH72}. 
%Parameters of calculations are given in the text.
Energies are given with respect to the $\alpha$+${^3{\rm He}}/{^3{\rm H}}$ threshold.}
\label{phaseshifts}
 \end{figure}

The $\alpha+{^3{\rm He}}$ and $\alpha+{^3{\rm H}}$ elastic phase shifts are computed for relative collision energies up to $\sim$10~MeV and shown in \Fig{phaseshifts}. For the sake of clarity, the jump of $+180^\circ$ in the phase shifts at the second $5/2^-$ and $7/2^-$ resonance energies are not displayed.
As a complementary information to the phase shifts, the scattering lengths for the $1/2^+$,  $1/2^-$, and $3/2^-$ partial waves are provided in \Tab{tab_scl}.
\begin{table}[h!]
\centering
\renewcommand{\arraystretch}{1.2}
{\small
\begin{tabular}{l r r r}
\hline
$\alpha+{^3{\rm He}}$ & NCSMC & ``Exp.''& Refs.\\
\hline
$a_{1/2^+}~[{\rm fm}]$&  7.7 & 41.06 & \cite{KB07}\\
$a_{1/2^-}~[{\rm fm}^3]$&263.9  & $413\pm 7 $ & \cite{YB11}\\
$a_{3/2^-}~[{\rm fm}^3]$&210.4  &  $301\pm 6$ & \cite{YB11}\\
\hline
$\alpha+{^3{\rm H}}$ & NCSMC & ``Exp.'' & Refs.\\
\hline
$a_{1/2^+}~[{\rm fm}]$&  5.4 & 13.05 & \cite{KB07}\\
$a_{1/2^-}~[{\rm fm}^3]$&82.6  & $95.13\pm 1.73 $ & \cite{YB11}\\
$a_{3/2^-}~[{\rm fm}^3]$& 70.0 &  $58.10\pm 0.65$ & \cite{YB11}\\
\hline
\end{tabular}
}
\caption{Scattering lengths $a_{J^\pi}$ associated with partial waves $J^\pi$ for the $\alpha+{^3{\rm He}}$ and $\alpha+{^3{\rm H}}$ collisions. Values infered in \Refs{KB07,YB11} from the experimental phase shifts are given for comparison in column "Exp" (see text for more details).}
\label{tab_scl}
\renewcommand{\arraystretch}{1}
\end{table}
For the $1/2^-$ and $3/2^-$ partial waves, they are compared with the values obtained in \Ref{YB11} from the experimental phase shifts and asymptotic normalization coefficients (ANC) by using relations linking the effective-range expansion and the ANC. Such a method is not applicable for the $1/2^+$ partial wave since there is no bound states in this partial wave. Therefore, for the $1/2^+$ partial wave, we cite the value obtained in \Ref{KB07} from a microscopic cluster model reproducing the experimental phase shits. This value cannot be considered, strictly speaking, as an experimental one but provides a reasonable point of comparison.

\begin{figure}
\centering
\scalebox{0.33}{\includegraphics{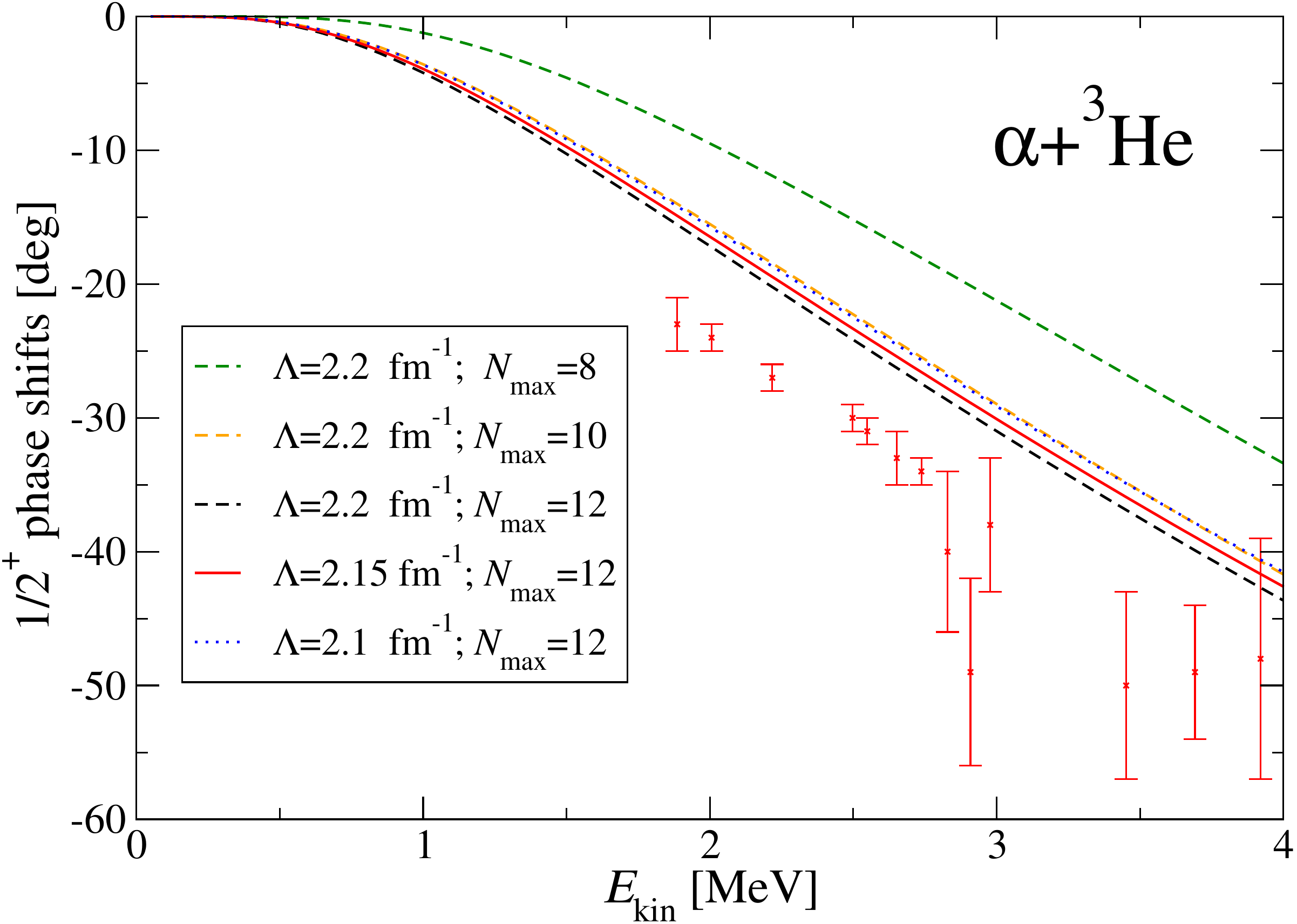}}
\caption{(Color online) $1/2^+$ phase shifts for different values of the SRG parameter: $\Lambda=2.1\ {\rm fm}^{-1}$ (dotted lines), $\Lambda=2.15\ {\rm fm}^{-1}$ (solid lines), and $\Lambda=2.2\ {\rm fm}^{-1}$ (dashed lines). 
For $\Lambda=2.2\ {\rm fm}^{-1}$, different values of $N_{\rm max}$ are considered; the $N_{\rm max}$ value used for computing the colliding-nuclei wave functions is given.}
\label{Swave}
 \end{figure}

In both systems, the $1/2^+$ theoretical phase shifts overestimate the corresponding experimental ones. 
To analyse this discrepancy, we focuses on one system, namely $\alpha+{^3{\rm He}}$. 
The $1/2^+$ phase shifts are displayed in \Fig{Swave} for three different values of the SRG parameter $\Lambda$ ($2.1$, $2.15$, and $2.2\ {\rm fm}^{-1}$) and the convergence with respect to $N_{\rm max}$ is illustrated for the harder potential  ($\Lambda=2.2\ {\rm fm}^{-1}$). The convergence for the two other values of $\Lambda$ is expected to be similar.
While the $1/2^+$ phase shifts are not fully converged at $N_{\rm max}=12$, the pattern of convergence indicates that, even by increasing $N_{\rm max}$, which is out of reach for computational reasons, the experimental phase shifts will not be reproduced for $\Lambda=2.2\ {\rm fm}^{-1}$. Neither will they be reproduced by considering the two other values of $\Lambda$ since the difference between the $1/2^+$ phase shifts for the three adopted values of $\Lambda$ is small. Based on these results, we can reasonably argue that the non-reproduction of the experimental  $1/2^+$ phase shifts by our approach is a feature of the two-nucleon forces used here and not a consequence of a non-fully converged calculation. Taking the three-nucleon forces into account could impact significantly the phase shifts.
The same conclusions can be drawn from the analysis of the $1/2^+$ scattering lengths given in \Tab{tab_as} for different values of the SRG parameter $\Lambda$ and different values of $N_{\rm max}$.
\begin{table}[h!]
\centering
{\small
\begin{tabular}{l r r}
\hline
$\Lambda\ [{\rm fm}^{-1}]$& $N_{\rm max}$ & $a_{1/2^+}$ [fm]\\
\hline
2.2 &  8 & -2.5 \\
2.2 &10 & 6.5\\
2.2 &12 & 9.1\\
2.15 &12 &7.7 \\
2.1 &12 & 6.2\\
\hline
\end{tabular}
}
\caption{$1/2^+$ scattering length for the $\alpha+{^3{\rm He}}$ collision for different values of the SRG parameter $\Lambda$ and different values of $N_{\rm max}$; the $N_{\rm max}$ value used for computing the colliding-nuclei wave functions is given.}
\label{tab_as}
\renewcommand{\arraystretch}{1}
\end{table}

\begin{figure}
\centering
\scalebox{0.37}{\includegraphics{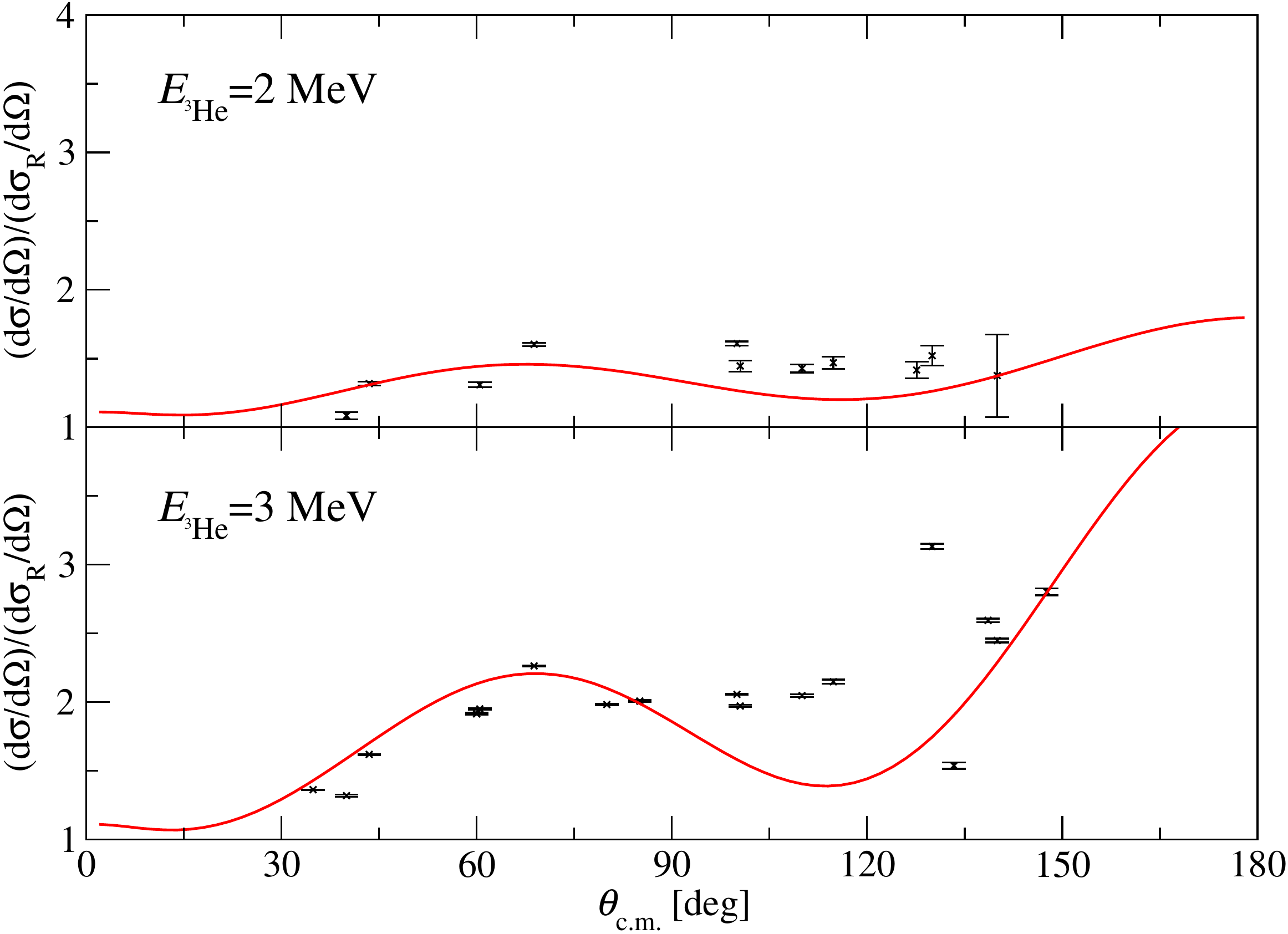}}
\caption{Differential $\alpha+{^3{\rm He}}$ elastic cross sections ($d\sigma/d\Omega$) normalized by the differential Rutherford cross sections ($d\sigma_R/d\Omega$) as a function of the scattering angle measured in the c.m.\ frame. Experimental data come from \Ref{MAZ93}.}
\label{fig_cs}
 \end{figure}

For negative-parity partial waves, the discrepancy between theoretical and experimental resonances seen in \Fig{spectra} is also visible in the phase shifts. Moreover, the splitting between the $1/2^-$ and $3/2^-$ is underestimated, as it can be seen from the comparison of the phase shifts and of the scattering lengths.
Instead of analysing the phase shifts and the scattering lengths, we can compare directly theoretical and experimental cross sections. In \Fig{fig_cs}, the differential $\alpha+{^3{\rm He}}$ elastic cross sections are displayed for different angles at two particular colliding energies and compared with experimental data from \Ref{MAZ93}, for which no phase-shift analysis exists. 
Our approach reproduces the general trends of the experimental data. 

To evaluate the impact of the discrepancies in the elastic scattering on the ${^3{\rm He}}(\alpha,\gamma){^7{\rm Be}}$ and ${^3{\rm H}}(\alpha,\gamma){^7{\rm Li}}$ astrophysical $S$ factors, we adopt a phenomenological model based on the NCSMC results in the largest model space. 
The basic idea is to consider the energies of the square-integrable NCSM basis states $E_\lambda$, appearing in \Eq{Elam}, as adjustable parameters. 
These new degrees of freedom are then used to reproduce the experimental ${^7{\rm Be}}$ and ${^7{\rm Li}}$ bound-state and resonance energies and reducing the gap between theoretical and experimental $1/2^+$ phase shifts.

\begin{figure}
\centering
 \scalebox{0.5}{\includegraphics{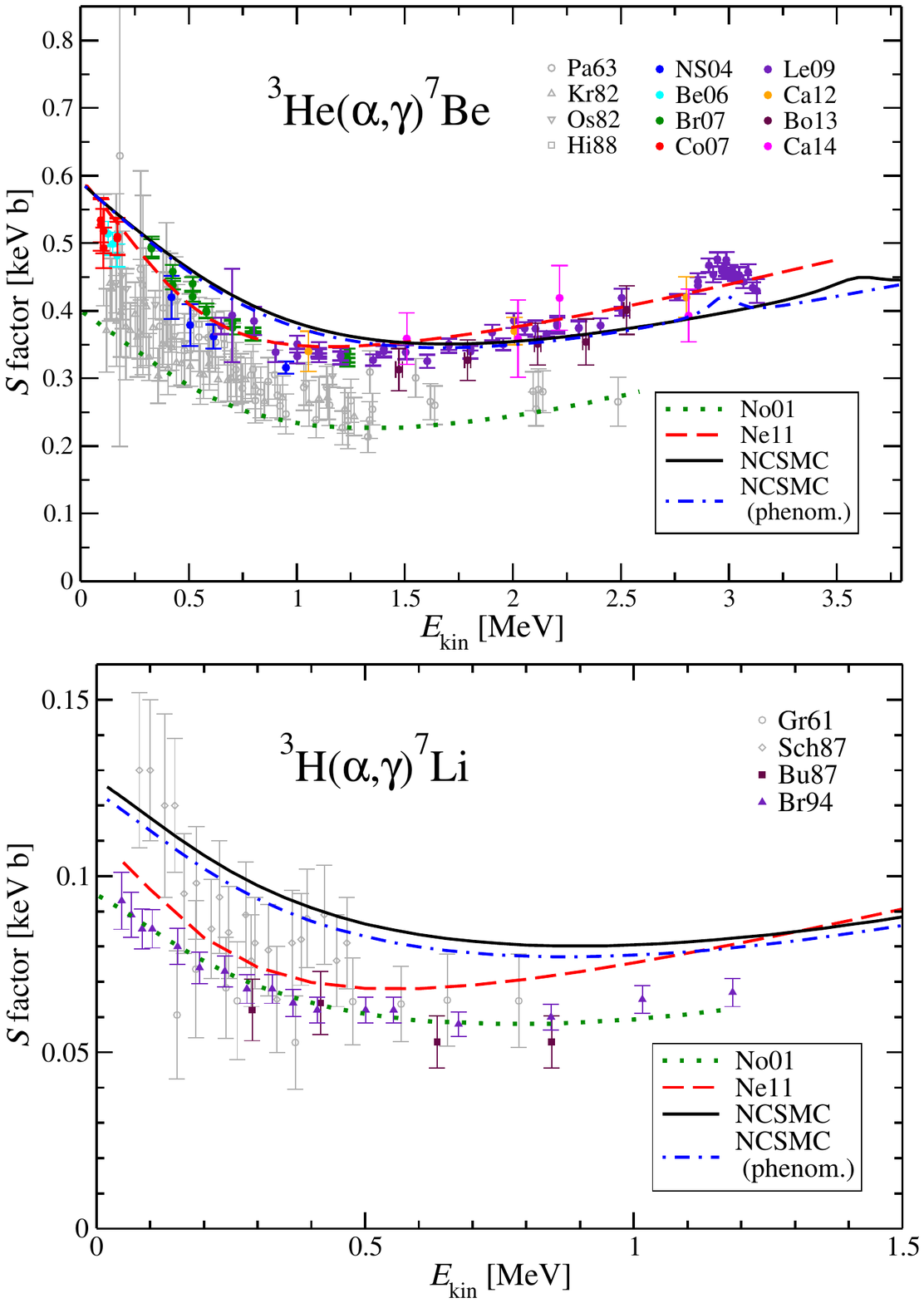}}
\caption{(Color online) Astrophysical $S$ factor for  the ${^3{\rm He}}(\alpha,\gamma){^7{\rm Be}}$ and ${^3{\rm H}}(\alpha,\gamma){^7{\rm Li}}$ radiative-capture processes obtained from the NCSMC approach and from its phenomenological version and compared with other theoretical approaches~\cite{No01,Ne11} and with experiments~\cite{Pa63,Kr82,Os82,Hi88,NHN04,BCC06,CBC07,BBS07,DGK09,CNB12,BGH13,Ca14,Gr61,Sch87,Bu87,BKR94}. Recent data are in color (online) and old data are in light grey.}
\label{Sfactor}
 \end{figure}

The ${^3{\rm He}}(\alpha,\gamma){^7{\rm Be}}$ and ${^3{\rm H}}(\alpha,\gamma){^7{\rm Li}}$ astrophysical $S$ factors obtained with the NCSMC approach and with its phenomenological version are displayed in \Fig{Sfactor} and compared with experiment~\cite{Pa63,Kr82,Os82,Hi88,NHN04,BCC06,CBC07,BBS07,DGK09,CNB12,BGH13,Ca14,Gr61,Sch87,Bu87,BKR94}.
The astrophysical $S$ factors extrapolated at zero colliding energy are given in \Tab{tab3}.
\begin{table}[h!]
\centering
\renewcommand{\arraystretch}{1.2}
\begin{tabular}{l l l l}
\hline
&NCSMC&Exp.&Refs.\\
\hline
$S_{{^3{\rm He}}(\alpha,\gamma){^7{\rm Be}}}(0)$ [keV b]&0.59&0.56(2)(2)&\cite{AGR11}\\
$S_{{^3{\rm H}}(\alpha,\gamma){^7{\rm Li}}}(0)$ [keV b]&0.13&0.1067(4)(60)&\cite{BKR94}\\
\hline
\end{tabular}
\caption{${^3{\rm He}}(\alpha,\gamma){^7{\rm Be}}$ and ${^3{\rm H}}(\alpha,\gamma){^7{\rm Li}}$ astrophysical $S$ factors extrapolated at zero collision energy. Experimental data come from \Refs{AGR11,BKR94}. For the ${^3{\rm He}}(\alpha,\gamma){^7{\rm Be}}$ reaction, the numbers in parentheses are the errors in the least significant digits coming from the experiments and from the theoretical extrapolation while for the ${^3{\rm H}}(\alpha,\gamma){^7{\rm Li}}$ reaction, they are the statistical and systematic errors.}
\label{tab3}
\renewcommand{\arraystretch}{1}
\end{table}
The electric $E1$ and $E2$ transitions as well as the magnetic $M1$ transitions have been considered. 
For the energy ranges which are considered, the contribution of the $E1$ transitions is dominant while $M1$ contribution is essentially negligible and the $E2$ transitions play a small but visible role in the ${^3{\rm He}}(\alpha,\gamma){^7{\rm Be}}$ radiative capture, mostly near the $7/2^-$ resonance energy.
Qualitatively, the ${^3{\rm He}}(\alpha,\gamma){^7{\rm Be}}$ astrophysical $S$ factors agree rather well with the experimental ones. 
The results obtained with the phenomenological model are similar up to approximately the $7/2^-$ resonance energy.
Indeed, the peak in the experimental $S$ factor at a relative collision energy of about 3~MeV corresponds to a $E2$ transition from the $7/2^-$ resonance to the $3/2^-$ ground state. 
Since the $7/2^-$ resonance energy is slightly overestimated by our theoretical approach, the energy of the corresponding peak in the $S$ factor is also overestimated. On the contrary, in the phenomenological approach, the experimental energy and width of the $7/2^-$ resonance are reproduced and hence the energy and width of the corresponding peak in the $S$ factor.
The adjustment of the ${^7{\rm Be}}$ square-integrable energies in the phenomenological approach to reproduce the experimental bound-state energies with an accuracy of $5$~keV or better has an impact of $1\%$ or less on the astrophysical $S$ factor for the energy range considered in \Fig{Sfactor}.
As it can be expected, the adjustment of the $5/2^-$ resonance energies has a negligible effect on the astrophysical $S$ factor, about $0.1\%$.
The $1/2^+$ phase shifts are not very sensitive to the energies of the square-integrable NCSM states, which are the adjustable parameters of our phenomenological model. Even by increasing these energies by amounts as large as 10~MeV, the $1/2^+$ phase shift is only reduced by $0.5^\circ$ at 1~MeV, $1.5^\circ$ at 2~MeV, and $2^\circ$ at 3~MeV. The consequent impact on the astrophysical $S$ factors is within few percents.
A further improvement of the ${^3{\rm He}}(\alpha,\gamma){^7{\rm Be}}$ astrophysical $S$ factors would require the inclusion of three-body forces and possibly the increase of the accuracy of the basis states, i.e., the increase of $N_{\rm max}$.
The importance of the three-body forces is also highlighted by comparing the results obtained with the NCSMC for different SRG resolution scales $\Lambda$ (not shown in the figures). Increasing (reducing) $\Lambda$ by 0.05~fm$^{-1}$ 
induces a reduction (an increase) of the astrophysical $S$ factor by an amount between about 30~eV~b and 60~eV~b over the energy range considered in  \Fig{Sfactor}.

\begin{figure}
\centering
\scalebox{0.37}{\includegraphics{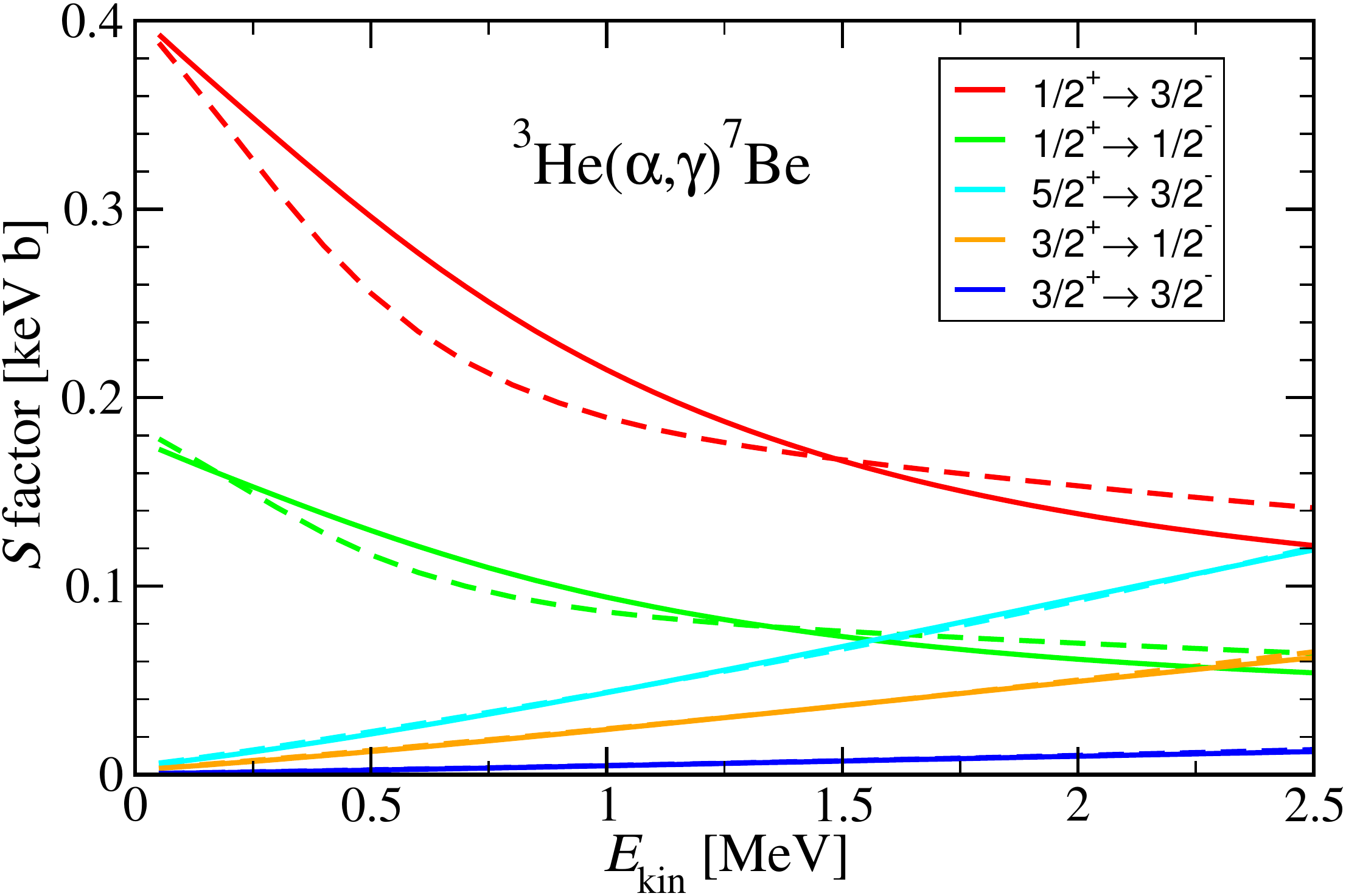}}
% \scalebox{0.37}{\includegraphics{figures/Be7contributions}}
\caption{(Color online) Partial-wave decomposition of the $E1$ contributions to the ${^3{\rm He}}(\alpha,\gamma){^7{\rm Be}}$ astrophysical $S$ factors with the NCSMC approach (solid lines) compared with Neff's calculations (dashed lines)~\cite{Ne12}.}
\label{decomp}
 \end{figure}

In \Fig{Sfactor}, other theoretical results based on two different realistic $NN$ interactions~\cite{No01,Ne11} are also presented.  
Nollet's approach~\cite{No01} is not fully microscopic but hybrid, based on both \textit{ab initio} variational Monte Carlo wave functions and phenomenological potential-model wave functions. 
In contrast, Neff's calculation~\cite{Ne11} is fully microscopic. As our approach, it is based on resonating-group method wave functions and the microscopic $R$-matrix to enforce the proper boundary conditions, but the model space is built from fermionic molecular dynamics (FMD) wave functions and only the $E1$ transitions are considered. 
Although the FMD approach is not fully able to describe the short-range correlations of the wave function, a good agreement between theoretical and experimental ${^3{\rm He}}(\alpha,\gamma){^7{\rm Be}}$ astrophysical $S$ factor was obtained ~\cite{Ne11}. 
Let us stress that Nollet's and Neff's approaches and the present one are -- for technical reasons -- based on three different $NN$ interactions. 
They are thus not supposed to give the same results and, in fact, both absolute values of the astrophysical $S$ factors and their 
energy dependence differ significantly.
To get some insights on these differences, the $E1$ contributions to the ${^3{\rm He}}(\alpha,\gamma){^7{\rm Be}}$ astrophysical $S$ factors obtained in this work are decomposed into the different partial waves and compared with Neff's results~\cite{Ne12} in \Fig{decomp}.
Because of the absence of the centrifugal barrier, the transitions from the $1/2^+$ partial wave ($\ell=0$) are the most important and those are the ones for which we find the largest difference.
A better knowledge of the empirical $\alpha+{^3{\rm He}}$ $s$-wave phase shifts could thus provide a useful test of the accuracy of the theoretical results.

The ${^3{\rm H}}(\alpha,\gamma){^7{\rm Li}}$ astrophysical $S$ factors are overestimated over the full energy range in our calculation. 
The phenomenological approach improves only slightly the situation.
As shown in \Fig{Sfactor}, a similar behavior is present, though less pronounced, in Neff's calculations~\cite{Ne11} while Nollet's approach~\cite{No01} reproduces the ${^3{\rm H}}(\alpha,\gamma){^7{\rm Li}}$ astrophysical $S$ factor but underestimates the  ${^3{\rm He}}(\alpha,\gamma){^7{\rm Be}}$ one. 
This suggests a possible underestimation of the experimental systematic uncertainties and underscores the need for new experimental studies of ${^3{\rm H}}(\alpha,\gamma){^7{\rm Li}}$ and a more complete microscopic calculation, including the effect of three-nucleon forces.
Again, the dependence of the results ons the SRG parameter has been studied. Increasing (reducing) of $\Lambda$ by 0.05~fm$^{-1}$ induces a reduction (an increase) of the astrophysical $S$ factor by about 10~eV~b over the energy range considered in \Fig{Sfactor}. It has to be noted that the calculations based on $\Lambda=2.15$~fm$^{-1}$ reproduce more accurately the seven-nucleon bound-state energies and therefore, the astophysical $S$ factors for this $\Lambda$ value should be more reliable.

\begin{figure}
\centering
\scalebox{0.4}{\includegraphics{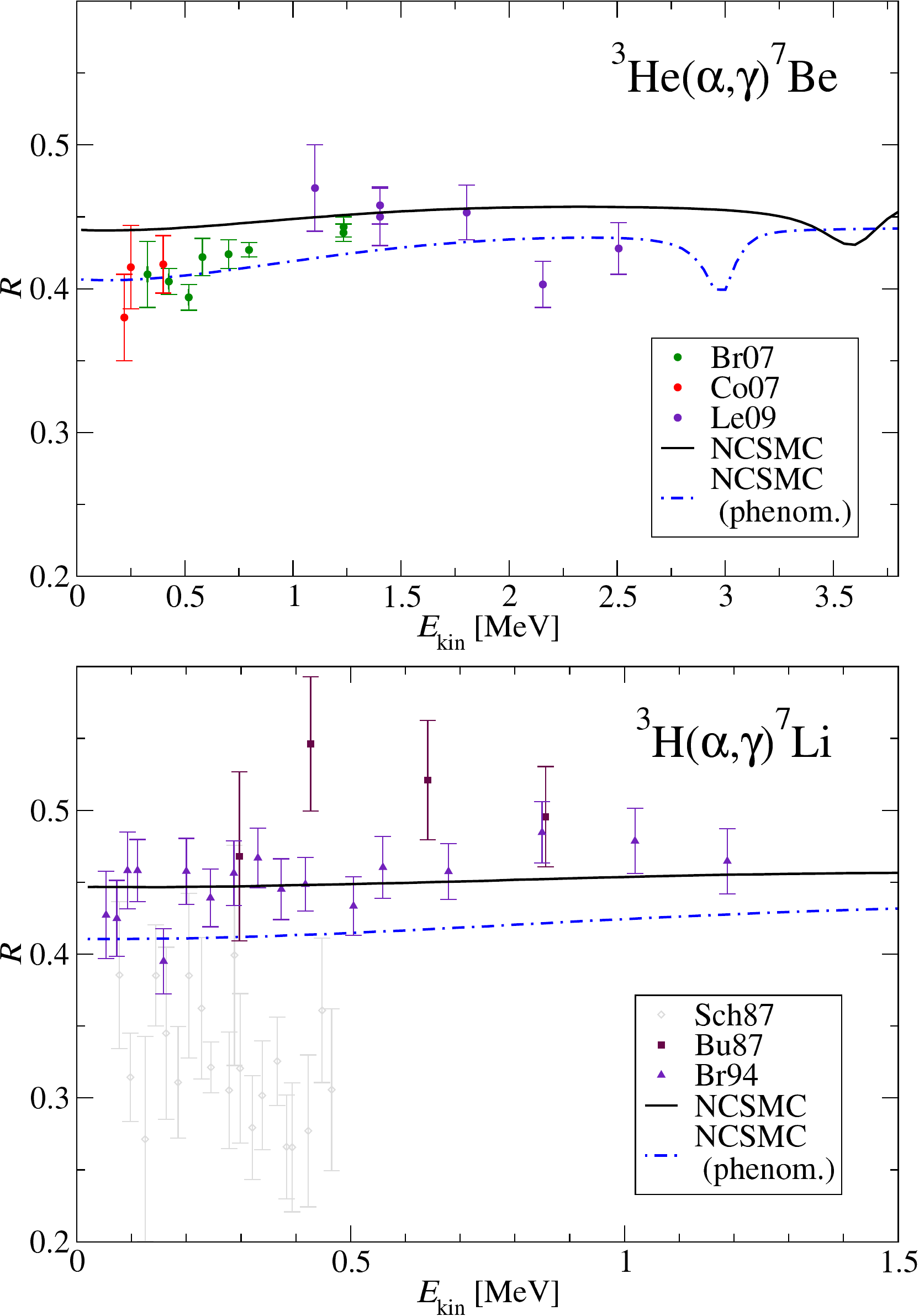}}
% \scalebox{0.25}{\includegraphics{ratioBe7}}
% \scalebox{0.25}{\includegraphics{ratioLi7}}
\caption{(Color online) Ratio ($R$) of the ${^3{\rm He}}(\alpha,\gamma){^7{\rm Be}}$  and ${^3{\rm H}}(\alpha,\gamma){^7{\rm Li}}$  radiative-capture cross sections to the seven-nucleon excited state and to the seven-nucleon ground state obtained from the NCSMC, from its phenomenological version and from experiments~\cite{BBS07,CBC07,DGK09,Sch87,Bu87,BKR94}.}
\label{ratio}
 \end{figure}

Finally, in \Fig{ratio}, we compare the ratio of the radiative-capture cross sections to the seven-nucleon excited state and to the seven-nucleon ground state obtained from our approaches and from experiments. Our theoretical results agree rather well with the experimental data. Differences between the NCSMC approach and its phenomenological version are comparable with the size of the experimental error bars.
\section{Conclusion}
In this letter, the ${^3{\rm He}}(\alpha,\gamma){^7{\rm Be}}$ and ${^3{\rm H}}(\alpha,\gamma){^7{\rm Li}}$ radiative-capture processes 
are described by means of the no-core shell model with continuum approach~\cite{BNQ13,BNQ13b}. 
Although the approach is restricted to two-nucleon forces, a rather good description of ${^7{\rm Be}}$ and  ${^7{\rm Li}}$ nuclei is obtained. Theoretical and experimental $\alpha+{^3{\rm He}}$ and  $\alpha+{^3{\rm H}}$ elastic phase shifts do not agree perfectly well. However, the discrepancy is difficult to characterize because of the lack of knowledge on the experimental uncertainties. New experimental studies of the $\alpha+{^3{\rm He}}$ and  $\alpha+{^3{\rm H}}$ elastic scattering would be highly desirable to probe more accurately the quality of the scattering wave functions. 
At low energies, the theoretical $s$-wave phase shifts are overestimated by our approach. This has a direct impact on the energy dependence of the astrophysical $S$ factor at low energies, which is therefore not exactly reproduced. The overestimation of the $s$-wave phase shifts is mostly due to the adjustment of the nucleon-nucleon interaction for reproducing the ${^7{\rm Be}}$ and ${^7{\rm Li}}$ binding energies without three-nucleon forces with the largest considered model space.
Taking the three-nucleon forces into account could thus reduce this discrepancy. Moreover, the inclusion of three-body forces would reduce the dependence of renormalization of the inter-nucleon interactions on the results, which would enable us to consider softer interactions, leading to a faster convergence. 
Including the three-body interactions is thus the next step in the development of an \textit{ab initio} approach of radiative-capture reactions. However,  this a particularly challenging task, which requires analytic developments and large computational efforts. Approximate ways to include the effects of three-body forces via the normal ordering technique~\cite{RBV12} are currently under study.
\section*{Acknowledgements}
We would like to thank Thomas Neff for useful discussions and, in particular, for sharing with us the results of his calculations and his extractions of experimental data.
We are also grateful to Barry Davids for fruitful discussions and for pointing out useful references to us. 
TRIUMF receives funding via a contribution through the National Research Council Canada. This work was
supported in part by NSERC under Grant No. 401945-2011, by LLNL under Contract DE-AC52-07NA27344, by the U.S. Department of Energy, Office of Science, Office of Nuclear Physics, under Work Proposal
Number SCW1158, and by JSPS KAKENHI Grant Numbers 25800121 and 15K05072. 
W.H. acknowledges Excellent Young Researcher Overseas Visit Program of JSPS
that allowed him to visit LLNL (2009-2010).
Computing support came from the LLNL institutional Computing Grand
Challenge Program and from an INCITE Award on the Titan supercomputer of the Oak
Ridge Leadership Computing Facility (OLCF) at ORNL.

\section*{References}
%\bibliography{../../../bib/biblio}

\newcommand{\bibGy}{Gy}\newcommand{\bibZs}{Zs}\newcommand{\bibPH}{P.-H}

\end{document}